\documentclass{ws-mpla}
\usepackage{lscape}
\begin{document}

\markboth{P.C. Srivastava}
{Shell model calculations in Ni region}
%%%%%%%%%%%%%%%%%%%%% Publisher's Area please ignore %%%%%%%%%%%%%%
\catchline{}{}{}{}{}
%%%%%%%%%%%%%%%%%%%%%%%%%%%%%%%%%%%%%%%%%%%%%%%%%%%%%%%%%%%%%%%%%%%

\title{ Nuclear structure study with core excitations in Ni region: for $fpg_{9/2}$
 space}

\author{P.C. SRIVASTAVA\footnote {praveen.srivastava@nucleares.unam.mx}}

\address{Department of Physics, University of Allahabad, Allahabad-211002, India,}
\address{Grand Acc\'el\'erateur National d'Ions Lourds (GANIL), CEA/DSM--CNRS/IN2P3, BP~55027, F-14076 Caen Cedex 5, France, and}
\address{Instituto de Ciencias Nucleares, Universidad Nacional Aut\'onoma de M\'exico, 04510 M\'exico, D.F., M\'exico } 

\maketitle

%\pub{Received (Day Month Year)}{Revised (Day Month Year)}

\begin{abstract}
Shell model calculations for Ni, Cu and Zn isotopes by modifying $fpg$
interaction due to Sorlin {\it et. al.,} [Phys. Rev. Lett.
88, 092501 (2002)] have been reported. In the present work 28 two body matrix elements of the
earlier interaction have been modified. Present interaction is able to
explain new experimental results for this region.

\keywords
{Large scale; Collectivity.}
\end{abstract}
\ccode{PACS 21.60.Cs; 27.50.+e}

%\end{frontmatter}
%=================================================================

\section {Introduction}
 \label{s_intro} 

The nickel isotopes ($Z=28$) cover three doubly-closed shells with number $N=28$, $N=40$, $N=50$ and therefore a unique testing ground to investigate the evolution of shell structure. 
The $^{68}$Ni and its neighboring attracted the interest of recent research to answer the magicity
versus superfluidity question related to doubly magic character of this nuclei. \cite{Bro95,Sor02,Lan03,Per06,Naka10} 
For the copper isotopes, the important question is related to rapid reduction in the energy of 5/2$^-$ state as the filling of neutrons started in the $\nu g_{9/2}$ orbital 
.\cite{Ste08,Ste09,Ily09,Dau10,Fla09,Fla10} In the Cu isotopes as $ N \sim 40$, the state reveal three types of structures -- single-particle, collective or coupling of single proton with neighboring even Ni isotopes. 
In the Zn isotopes with two protons more than the semi-magic $Z=28$ nickel nuclei, the spherical $N=40$ gap may not be strong enough to stabilize the nuclei in spherical shape when protons are added to the $^{68}$Ni core. Thus $B(E2)$ value is larger in
the chain of Zn isotopes. To understand the evolution of nuclear structure, the importance of monopole term from the tensor force is pointed out by Otsuka {\it et al.} \cite{Ots05,Ots10}

\begin{figure}[h]
\begin{center}
\resizebox{95mm}{!}{\includegraphics{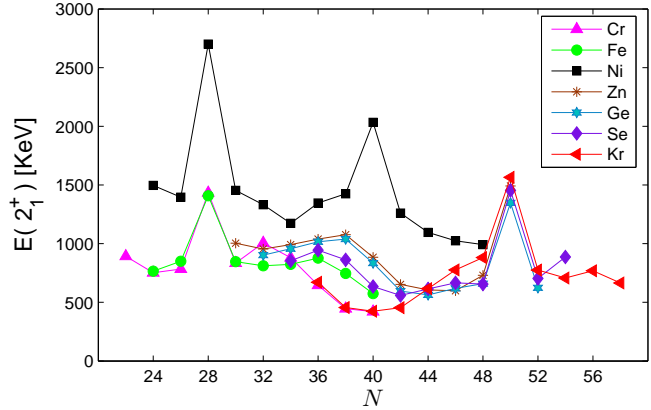}}
\end{center}
\caption{Systematic of the experimentally observed $E(2_1^+)$ for $Z=28$ to $Z=36$ near the $N~=~28,~40$ and $~50$ shell closure.}
\end{figure}

The experimental $E(2_1^+)$ for $Z=28$ to $Z=36$ are shown in Fig. 1. In this figure $E(2_1^+)$ states in Zn are overall lower compared to Ni and an additional decrease of the $E(2_1^+)$ energy in Zn isotopes is obtained between $N$=40-50 compared to $N$=28-40. This figure also reveal the enhancement of collectivity beyond $Z=30$, because of rapid decrease of $E(2^+$).
Below the Ni chain, the spectroscopy of Cr, Fe isotopes shown an increased collectivity toward
$N=40$, revealing the collapse of this shell closure. \cite{Gade10,Roth11}

Following our recent shell-model (SM) studies for
neutron-rich F isotopes \cite{Sri11}, odd and even isotopes of Fe
\cite{Sri09,Sria}, odd-odd Mn isotopes \cite{Srib},
odd-mass $^{61,63,65}$Co isotopes \cite{Sric}, and odd-even $^{71-81}$Ga isotopes \cite{Sric1}, in the present work, large scale shell model calculations have been performed for neutron rich Ni, Cu and Zn isotopes for $40 \leq N \leq50$ in $fpg_{9/2}$ model space. The low-lying energy levels and $B(E2)$ values have been calculated and compared with the recent experimental data.

The paper is organized as follows: In Section 2 details of calculation are described.
In Section 3 results and discussion are presented. Finally in Section 4 we give conclusions.

%=================================================================

\section{ Details of Calculation}

Large scale shell model calculations have been performed for neutron rich
nickel, copper and zinc isotopes with 40$\leq$N$\leq$50 using $^{40}$Ca as a
core by including the $f_{7/2}$ orbit. In the previous work \cite{Sri-th},
large scale shell model calculations were performed for neutron rich nickel,
copper and zinc isotopes for 40$ \leq$ N $\leq$50, using three different
versions \cite{Jen95,Now96,Smi04,Lis04,Honma09} of effective interactions for the model
space consisting of the $p_{3/2}$, $p_{1/2}$, $f_{5/2}$ and $g_{9/2}$ . Both,
however, yield unsatisfactory results in certain aspects, viz.\\
(i)  large E($2^+$) value for very neutron rich nuclei ($^{76}$Ni and
 $^{80}$Zn),\\
 (ii) small $B(E2)$ values in comparison to experimental
 values, and\\
 (iii) for $^{75}$Cu, $^{77}$Cu and $^{79}$Cu, the ground state
 is 3/2$^-$ as compared to the experimental indication of 5/2$^-$.\\ 

 This is probably due to neglect of the  $f_{7/2}$ orbit for the protons. In
view of this, $f_{7/2}$ orbit has been included in the valence space of
protons by taking $^{40}$Ca as core. Effective interaction for $fpg$ valence
space for both protons and neutrons with $^{40}$Ca as core has been
constructed by Sorlin $\it {et~ al.}$ \cite{Sor02} The main drawback of this
interaction is that the effective proton single-particle energies of
$f_{7/2}$ orbital become lower than $f_{5/2}$ which is not realistic. In the
present work, we have modified relevant matrix elements to account for
this discrepancy.

\subsection{Model space}   

In the present work, we have used $fpg$ model space comprising of the 
$0f_{7/2}$, $1p_{3/2}$, $0f_{5/2}$, $1p_{1/2}$ active proton orbitals and
$0f_{7/2}$, $1p_{3/2}$, $0f_{5/2}$, $1p_{1/2}$, $0g_{9/2}$ neutron orbitals
with eight $f_{7/2}$ frozen neutrons, more accurately $^{40}$Ca core with
eight f$_{7/2}$ frozen neutrons. The single-particle energies for $0f_{7/2}$,
$1p_{3/2}$, $0f_{5/2}$, $1p_{1/2}$, $0g_{9/2}$ orbitals are 0.0, 2.0, 6.5, 4.0
and 9.0 MeV respectively. The relative single-particle energies are taken
from the excitation energies of the low-lying negative parity states in
$^{49}$Ca. \cite{Sor02}

\subsection{Effective Interaction} 

For the $fpg$ valence space, an effective interaction  with  $^{40}$Ca
as core has been reported in Ref. \cite{Sor02} This interaction has been
built using $\it{fp}$ two-body matrix elements (TBME) from Ref. \cite{Pov01}
and $1p_{3/2}$, $0f_{5/2}$, $1p_{1/2}$, $0g_{9/2}$ TBME from Ref.
\cite{Now96} For the common active orbitals in these subspaces, matrix
elements were taken from. \cite{Now96} As the latter interaction has been
defined for a $^{56}$Ni core, a scaling factor of A$^{-1/3}$ amplitude is
applied to take into account the change of radius between the $^{40}$Ca and
$^{56}$Ni cores. As  more and more neutrons are added in the $f_{7/2}$ shell,
the excitation energy of the 9/2$^+$ state decreases
% due to the attractive $\pi f_{7/2}\nu g_{9/2}$ interaction
 as reflected in Fig. 2 for $^{41-71}$Sc
isotopes. The remaining matrix elements are taken from $f_{7/2}$$g_{9/2}$ 
TBME from Ref. \cite{Kah69}

\begin{figure}[h]
\begin{center}
\resizebox{120mm}{!}{\includegraphics{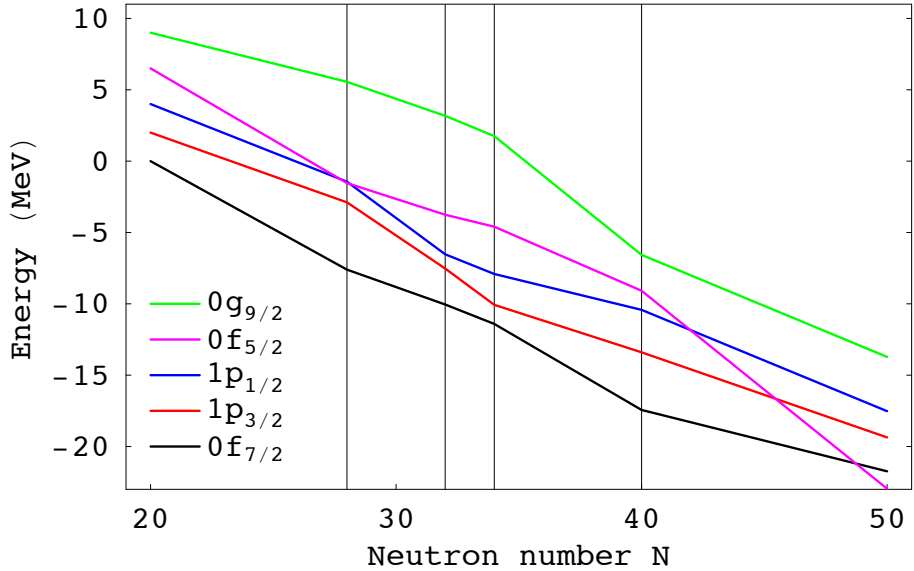}}
\end{center}
\caption{Effective single-particle energies in $^{41-71}$Sc isotopes using $\it {pfg9}$  interaction.}
\end{figure}

\subsection{Monopole correction to the single-particle levels}  

As more and more neutrons are added in $g_{9/2}$ orbital, $f_{7/2}$ orbital
becomes more repulsive and $f_{5/2}$ more attractive. Thus proton $f_{5/2}$
is pulled down while $f_{7/2}$ is lifted up, as $N$ increases, this is due to
monopole interaction produced by tensor force between a proton in j$_{>,<}$ =
${\it l}$ $\pm$1/2 and a neutron in j$^\prime_{>,<}$ = ${\it l}$
$^\prime$$\pm$1/2. Effects on $p_{3/2}$ and $p_{1/2}$  orbitals are small and
can be neglected. In the present work, we have  modified  $g_{9/2}$$f_{7/2}$
matrix elements by subtracting 100 keV and $g_{9/2}$$f_{5/2}$ matrix elements
by adding 100 keV. We have also check other set of values like 50 and 150 keV
to modify $g_{9/2}$$f_{7/2}$ and $g_{9/2}$$f_{5/2}$ set of matrix elements. 
The results with 100 keV modification is more reasonable.
 This modified interaction is named as $\it {pfg9a}$.
The effective single-particle energies for $^{41-71}$Sc isotopes are shown in Fig.3.

\begin{figure}[h]
\begin{center}
\resizebox{135mm}{!}{\includegraphics{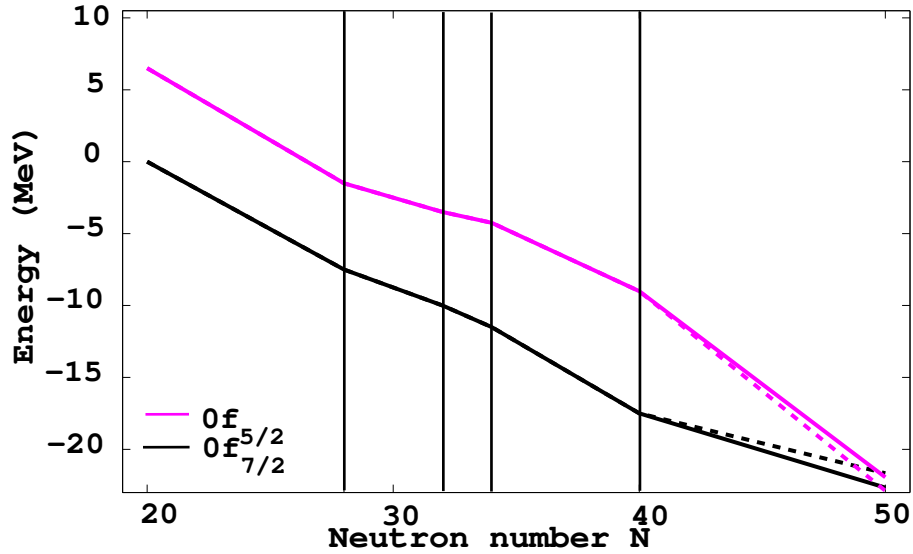}}
\end{center}
\caption{Effective single-particle energies in $^{41-71}$Sc isotopes using $\it {pfg9a}$  interaction.  In this figure,  the results for $pfg9$
interaction  are also shown as dotted lines.}
\end{figure}

\subsection{State-of-the-art in the calculations} 

The modified interaction has been modified to reproduce the experimental data of
Cu isotopes. These being odd nuclei, the available experimental data is
sparse. The aim of this tuning is to first reproduce the ground state
properties of $^{75-79}$Cu and then apply it to Ni and Zn isotopes. In the
original $fpg$ interaction, $f_{5/2}$  level crosses the $f_{7/2}$ at $^{70}$Sc
and become lower in energy. This appears to be unrealistic.

%\begin{figure}
%\begin{center}
%\resizebox{90mm}{!}{\includegraphics{Fig4.eps}}
%\caption{ The TBME as a function of J for pfg9 interaction.}
%\end{center} 
%\end{figure} 

%\begin{figure}
%\begin{center}
%\resizebox{90mm}{!}{\includegraphics{Fig5.eps}}
%\caption{ The changed TBME as a function of J for pfg9a interaction.}
%\end{center}
%\end{figure} 

\begin{figure} 
\begin{center}
\resizebox{50mm}{!}{\includegraphics{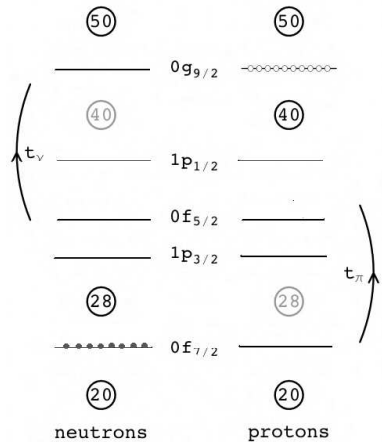}}
\end{center} 
\caption{Model space and truncation,
neutron: t$_\nu$ jumps from ($p_{3/2}$,$f_{5/2}$,$p_{1/2}$) to $g_{9/2}$, protons: $t_\pi$ jumps from $f_{7/2}$ to ($p_{3/2}$,$f_{5/2}$,$p_{1/2}$) orbital.}
\end{figure} 

In this modified interaction the $f_{7/2}$ effective single-particle energy
is always lower than $f_{5/2}$ which is expected. 
%The original TBME of $\it
%{pfg9a}$  interaction are shown in Fig. 4 and changed TBME for new
%interaction $\it {pfg9a}$ are shown in Fig. 5.
 Since the dimension of
matrices involved for $pfg9/2$ space is very large, truncation of the full
shell-model space are necessary. We have allowed neutron: $t_\nu$ jumps from
($p_{3/2}$, $f_{5/2}$, $p_{1/2}$) to $g_{9/2}$, protons: $t_\pi$ jumps from
$f_{7/2}$ to ($p_{3/2}$, $f_{5/2}$, $p_{1/2}$) orbital for each nucleus and
each isotopes, an example of this truncation are shown in Fig. 4. 
We performed calculations for maximum dimension  which are feasible by SGI-Cluster computer at GANIL. Up to this dimension, the states seems to be  more convergent,
which can be taken as the final result.

%For Cu, Ni and Zn nuclei the dimensions for different
%isotopes in m-scheme are tabulated in Table 1, 2 and 3. The highest
%dimensions which are feasible by SGI-cluster computer at GANIL are shown by
%bold face. Up to this dimension, the states seems to be  more convergent,
%which can be taken as the final result.

All the SM calculations have been carried out using the
code \textsc{antoine} ~\cite{Now96,Caurier05,Caurier99} at SGI-Cluster computer at GANIL
and DGCTIC-UNAM computational facility KanBalam.

\subsection{Binding Energies}

To compare our shell model results with the experimental binding energies relative to binding energies of $^{40}$Ca
we use following formula:

\begin{equation}
E_{\rm B} = -B_{\rm e}= E(SM)+E_{\rm C}+E_{\rm M}.
\end{equation}
In this Coulomb energies relative to $^{40}$Ca given by \cite{cole37}

\begin{equation}
E_{\rm C} =  e_{\pi} \pi + V_{\pi \pi} \frac{\pi(\pi -1)}{2} + V_{\pi \nu} \pi \nu
 + [\frac{1}{2}\pi]b_{c},
\end{equation}
where $\pi$($\nu$) stands for the number of valence protons (neutrons) for $^{40}$Ca core.
We have taken the values of parameters $e_{\pi}$, $V_{\pi \pi}$,
$V_{\pi \nu}$ and  $b_{c}$ from Ref.\cite{cole37} These parameters are determined by fitting to the measured Coulomb displacement energies
of Ni(Z=28)$\leq Z \leq$ Mo(Z=42) with 32$\leq N \leq$ 50. The resulting parameters are
\begin{equation}
e_{\pi}= 9.504~{\rm MeV},\\
V_{\pi \pi}=0.228~{\rm MeV},\\
V_{\pi \nu}=-0.036~{\rm MeV},\\
b_{c}=0.030~{\rm MeV}.
\end{equation}

The monopole expression is given by  \cite{Caurier99a}
\begin{equation}
E_{\rm M} =  e_{\nu}n+a \frac{1}{2}n(n-1)+b(T(T+1)-\frac{3}{4}n).
\end{equation}
where $n$ is total number of valence particles and $T$ is the total isospin. The 
$e_{\nu}$ is an average particle core interaction and $a$ and $b$ are the isoscalar and isovector global monopole correction.
The values of the parameters at $A=42$ are  \cite{Caurier99a}
\begin{equation}
e_{\nu}= -861 \pm 0.01~{\rm MeV},\\
a=0.041 \pm 0.003~ {\rm MeV},\\
b=0.119\pm 0.006~ {\rm MeV}.
\end{equation}

In Table 1, we have tabulated experimental (Expt.) and shell model (SM) g.s. energies.
%\newpage
\begin{table}
\tbl{The experimental (Expt.) and shell model (SM) g.s. energies. The values of energies are in MeV.}
{\begin{tabular}{@{}c|c|c|c|c|c@{}} 
 \toprule
           $^{68}$Ni$_{40}$  & $^{70}$Ni$_{42}$ & $^{72}$Ni$_{44}$     & $^{74}$Ni$_{46}$       &~ $^{76}$Ni$_{48}$ & \\	   
\hline
             Expt. ~~SM  &    Expt. ~~SM     &   Expt. ~~SM     &   Expt. ~~SM    &  Expt. ~~SM   & \\
\hline
             -248.3 ~ -227.7  &  -260.2 ~-239.5  & -271.1~ -249.9 & -281.7 ~-259.5 &  -293.7 ~ -267.3 & \\
\hline
          $^{69}$Cu$_{40}$  & $^{71}$Cu$_{42}$ & $^{73}$Cu$_{44}$     & $^{75}$Cu$_{46}$       &~ $^{77}$Cu$_{48}$     &~ $^{79}$Cu$_{50}$\\   
\hline
             Expt.~~SM    &    Expt. ~~SM   &   Expt. ~~SM   &   Expt. ~~SM  &  Expt.~~SM  &  Expt. ~~SM  \\
\hline
             -257.9 ~ -235.1  &  -271.0~ -248.2  & -283.4~ -259.7 & -294.7~ -270.4 &  -305.4~  -279.4  &  -315.2~  -287.2 \\
\hline
          $^{70}$Zn$_{40}$  & $^{72}$Zn$_{42}$ & $^{74}$Zn$_{44}$     & $^{76}$Zn$_{46}$       &~ $^{78}$Zn$_{48}$     &~ $^{80}$Zn$_{50}$\\	   
\hline
             Expt. ~~SM    &    Expt. ~~SM  &   Expt. ~~SM  &   Expt. ~~SM  &  Expt. ~~SM  &  Expt. ~~SM  \\
\hline
             -269.0~  -242.6  &  -283.7~ -257.4  & -297.5~ -271.8 & -310.0~ -283.6 &  -321.3~  -293.5  &  -332.0~  -302.4 \\
\botrule 
\end{tabular}\label{t_bee22}  }
\end{table}

%=================================================================

\section{ Results and Discussion}

The yrast levels for the  nickel, copper and zinc isotopes for 40$\le$ N $\le$
50  are shown in Figs.5, 6 and 7.

\subsection{Copper isotopes}

For $^{69}$Cu, the ground state 3/2$^-$ is well predicted by $pfg9a$ 
interaction. The calculated 1/2$^-$ and 5/2$^-$ state is reverse in comparison
to experimental levels. The calculated 1/2$^-$ and 5/2$^-$ state is 134 keV and
378 keV lower in energy from the experimental levels. The order of 7/2$_{1}^-$,
7/2$_{2}^-$ and 9/2$^-$ is well reproduced by this interaction. The  7/2$_{2}^-$
is only 62 keV below than experimental value. In $^{69}$Cu, the  3/2$^-$ is
interpreted by $\pi$p$_{3/2}$, the 5/2$^-$ is interpreted by $\pi$p$_{5/2}$. The
first 7/2$^-$ is interpreted by $\pi$$p_{3/2}$$\otimes$2$^+$($^{68}$Ni) and the
second 7/2$^-$ is interpreted by $\pi$$p_{7/2}^{-1}$.\\

\begin{figure}
\begin{center}
\resizebox{120mm}{50mm}{\includegraphics{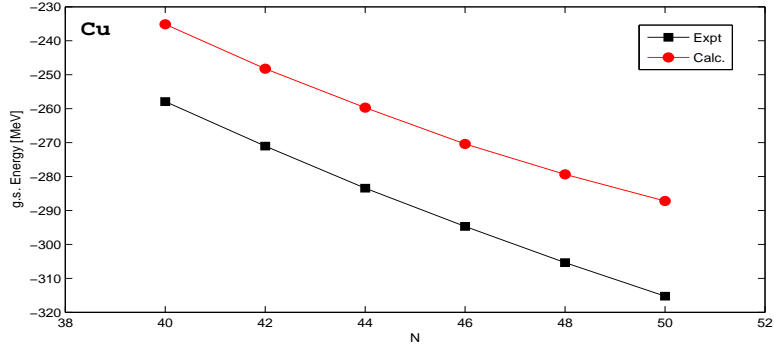}}
\caption{Experimental \protect\cite{audi03} and calculated ground-state energies for Cu isotopes from A=69 to 79.}
\end{center}
\end{figure}

\begin{figure}[h]
\begin{center}
\resizebox{120mm}{!}{\includegraphics{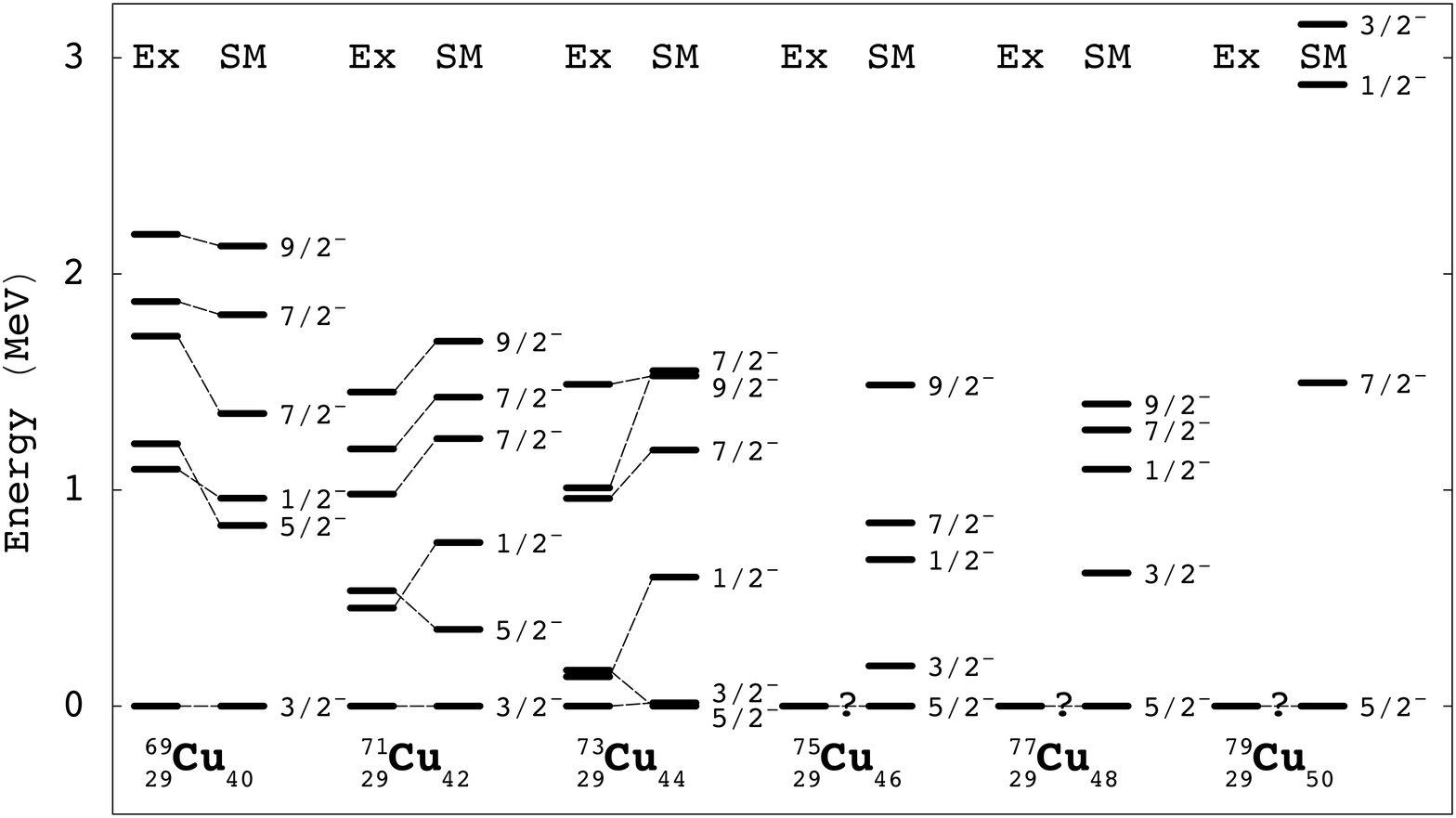}}
\end{center}
\caption{Yrast levels of $^{69-79}$Cu isotopes with $pfg9a$ interaction.}
\end{figure}
The experimental and calculated ground-state energies for Cu isotopes are shown in Fig. 5.
In Fig. 6, the yrast levels of $^{69-79}$Cu isotopes with $pfg9a$ interaction are shown.
For $^{71}$Cu, the ground state 3/2$^-$ is well predicted by $pfg9a$
interaction. The order of 5/2$^-$ and 1/2$^-$ state is reverse in comparison to
experimental levels. The calculated 1/2$^-$ level is 303 keV higher than the
experimental value, while 5/2$^-$ level is 179 keV lower than the experimental
result. The order of 7/2$_{1}^-$, 7/2$_{2}^-$ and 9/2$^-$ is well reproduced and
are higher in energy from the experimental values. In the $^{71}$Cu, the 
3/2$^-$ is interpreted by $\pi$$p_{3/2}$, and the 5/2$^-$ is interpreted by
$\pi$$p_{5/2}$. The first 7/2$^-$ is interpreted by
$\pi$$p_{3/2}$$\otimes$2$^+$($^{70}$Ni) and the second 7/2$^-$ is interpreted by
$\pi$$p_{7/2}^{-1}$.\\

For $^{73}$Cu, $pfg9a$ interaction predicts 5/2$^-$ state lower in energy
by 15 keV from  3/2$^-$ state, while experimentally 3/2$^-$ is ground state.
1/2$^-$ state is 462 keV higher in energy from experimental value. The 9/2$^-$
lies in between two 7/2$^-$ states. In the $^{73}$Cu, the  3/2$^-$ is
interpreted by $\pi$$p_{3/2}$, and the 5/2$^-$ is interpreted by $\pi$$p_{5/2}$.
The first 7/2$^-$ is interpreted by $\pi$$p_{3/2}$$\otimes$2$^+$($^{72}$Ni) and
the second 7/2$^-$ is interpreted by $\pi$$p_{7/2}^{-1}$.\\

Recently, there is experimental indication of 5/2$^-$ as a ground state in
$^{75}$Cu at REX-ISOLDE, CERN. \cite{Fla09} The $pfg9a$ interaction also
predicts 5/2$^-$ as a ground state. For $^{77}$Cu and $^{79}$Cu, this
interaction predicts 5/2$^-$ as a ground state which is also expected from
experiment. The calculations predict that the 1/2$^-$, 9/2$^-$ and 7/2$^-$
states are at very high in energy for $^{79}$Cu isotopes.

\subsection{Nickel isotopes} 

The nickel isotopes ($Z=28$) cover three doubly-closed shells and therefore,
 a unique testing ground for large scale shell model
calculations.  Experimentally, the 8$^+$ isomerism though expected to be present
in a whole chain from $^{72}$Ni to $^{76}$Ni was found to be suddenly absent in
$^{72}$Ni and $^{74}$Ni and present in $^{70}$Ni and $^{76}$Ni. The ground state
spin from $^{68}$Ni to $^{76}$Ni is correctly reproduced by $pfg9a$
interaction. 

\begin{figure}
\begin{center}
\resizebox{120mm}{50mm}{\includegraphics{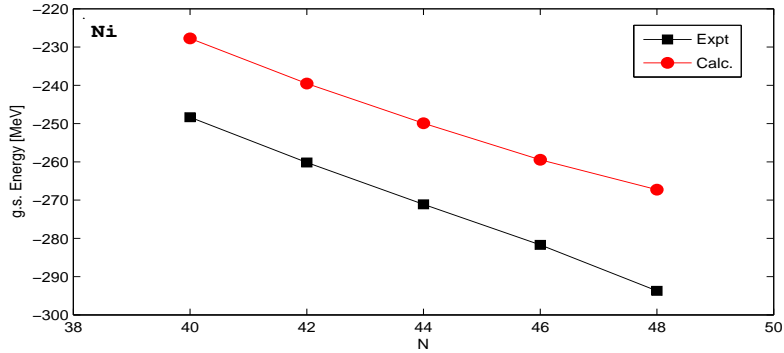}}
\caption{Experimental \protect\cite{audi03} and calculated ground-state energies for Ni isotopes from A=68 to 76.}
\end{center}
\end{figure}

\begin{figure}
\begin{center}
\resizebox{125mm}{60mm}{\includegraphics{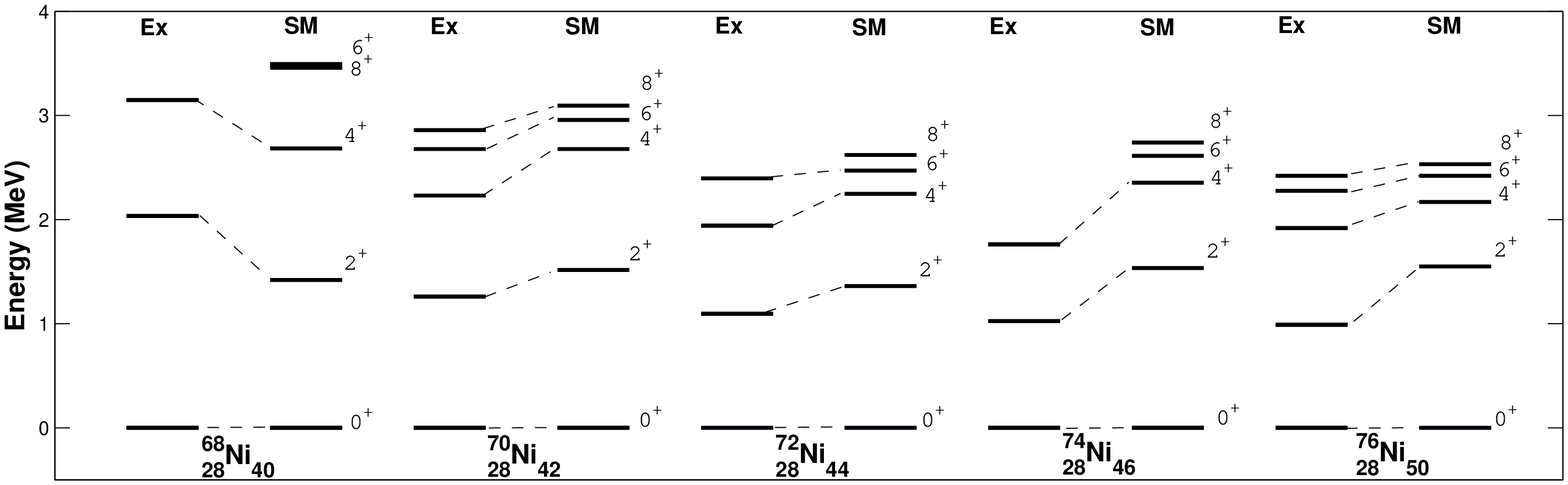}}
\caption{Yrast levels of $^{68-76}$Ni isotopes with $pfg9a$ interaction.}
\end{center}
\end{figure}

The experimental and calculated ground-state energies for Ni isotopes are shown in Fig. 7.
In Fig. 8, the yrast levels of $^{68-76}$Ni isotopes with $pfg9a$ interaction are shown.
For $^{70}$Ni, the first excited state 2$^+$ is calculated at 1.516 MeV which is
257 keV above than the experimental level. The states 4$^+$ at 2.229 MeV, 6$^+$ at
2.678 MeV, and 8$^+$ at 2.860 MeV are well predicted at 2.677 MeV, 2.958 MeV and
3.095 MeV. The agreement with the experimental data is reasonable. For $^{72}$Ni 
the first excited state 2$^+$ is calculated at 1.362 MeV which is 266 keV above
the experimental level. The states 4$^+$ at 1.941 MeV and 6$^+$ at 2.396 MeV, are
well predicted at 2.248 MeV and 2.470 MeV. Theoretically, there is 8$^+$ state at
2.621 MeV, but experimentally their is no indication of 8$^+$ isomeric states.
For $^{74}$Ni, only two excited states are experimentally known. The first
excited state  2$^+$ is calculated at 1.534 MeV which is 510 keV above the
experimental value. The state 4$^+$ at 1.763 MeV is predicted at 2.354 MeV. The
calculated 6$^+$, 8$^+$  state is at 2.612 MeV and 2.740 MeV respectively. For
$^{76}$Ni, the calculated first 2$^+$ state is slightly higher in energy about
557 keV from the experimental value. The states 4$^+$ at 1.922 MeV, 6$^+$ at 2.276
MeV and isomeric states 8$^+$ at 2.420 MeV are well predicted at 2.169 MeV, 2.420
MeV and 2.532 MeV. The agreement for 4$^+$, 6$^+$ and 8$^+$ states with the
experiment is good.

\subsection{Zinc isotopes} 
The experimental and calculated ground-state energies for Ni isotopes are shown in Fig. 9.
In Fig. 10, the yrast levels of $^{70-80}$Zn isotopes with $pfg9a$ interaction are shown.
The ground state spin for $^{70-80}$Zn isotopes for 40$\leq$N$\leq$50 is
correctly reproduced by $pfg9a$ interaction. \\ For $^{70}$Zn, the first
excited 2$^+$ state is calculated at 0.603 MeV which is 282 keV below than
experimental value. The states 4$^+$ at 1.79 MeV, 6$^+$ at 2.89 MeV and 8$^+$ at
3.75 MeV are well predicted at 1.47 MeV, 2.37 MeV and 2.53 MeV. The calculated
values are compressed in comparison to the experimental value.  For $^{72}$Zn,
experimentally only 2$^+$ state at 0.653 MeV is known, which is predicted well
by $pfg9a$ interaction with energy difference of 74 keV. The calculated
values of states 4$^+$, 6$^+$ and 8$^+$ are at 1.178, 1.923 and 2.601 MeV
respectively. For $^{74}$Zn, the first excited 2$^+$ state is calculated at
0.825 MeV which is 219 keV above the experimental value. The state 4$^+$ at
1.419 MeV is predicted at 1.551 MeV. The states 6$^+$ and 8$^+$ are at 2.783 MeV
and 3.430 MeV. For $^{76}$Zn, the first excited 2$^+$ state is calculated at
0.997 MeV which is 398 keV above the experimental value. The first 4$^+$ state
is calculated with energy difference of 679 keV from the experimental value. The
calculated 4$^+$, 6$^+$ and 8$^+$ states are at 1.975 MeV, 3.036 MeV and 3.243
MeV respectively.  For $^{78}$Zn, the first excited 2$^+$ state is calculated at
1.307 MeV, which is 477 keV above the experimental value. The states 4$^+$ at
1.621 MeV, 6$^+$ at 2.528 MeV and 8$^+$ at 2.673 MeV are predicted at 2.247 MeV,
2.923 MeV and 2.955 MeV. Experimentally, the difference between 6$^+$ and 8$^+$
states is 145 keV while theoretically, it is only 32 keV. For $^{78}$Zn, the
first excited 2$^+$ state is calculated at 1.953 MeV which is 461 keV smaller
than the experimental value.\\ The calculated and experimental E(2$_1^+$) and
E(4$_1^+$) for Ni and Zn isotopes are shown in Fig. 11. The high values of 
E(2$_1^+$) for Ni at N=40 is an indication of shell closure at N=40. For Zn
isotopes, the E(2$_1^+$) and E(4$_1^+$) at N=40 and 50 are high in comparison to
neighboring isotopes  reflects shell closure at N=40 and N=50.

\begin{figure}
\begin{center}
\resizebox{120mm}{50mm}{\includegraphics{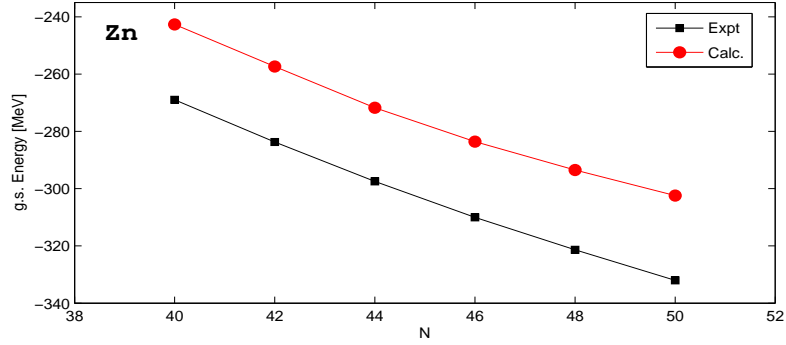}}
\caption{Experimental \protect\cite{audi03} and calculated ground-state energies for Zn isotopes from A=70 to 80.}
\end{center}
\end{figure}

\begin{figure}[h]
\begin{center}
\resizebox{135mm}{60mm}{\includegraphics{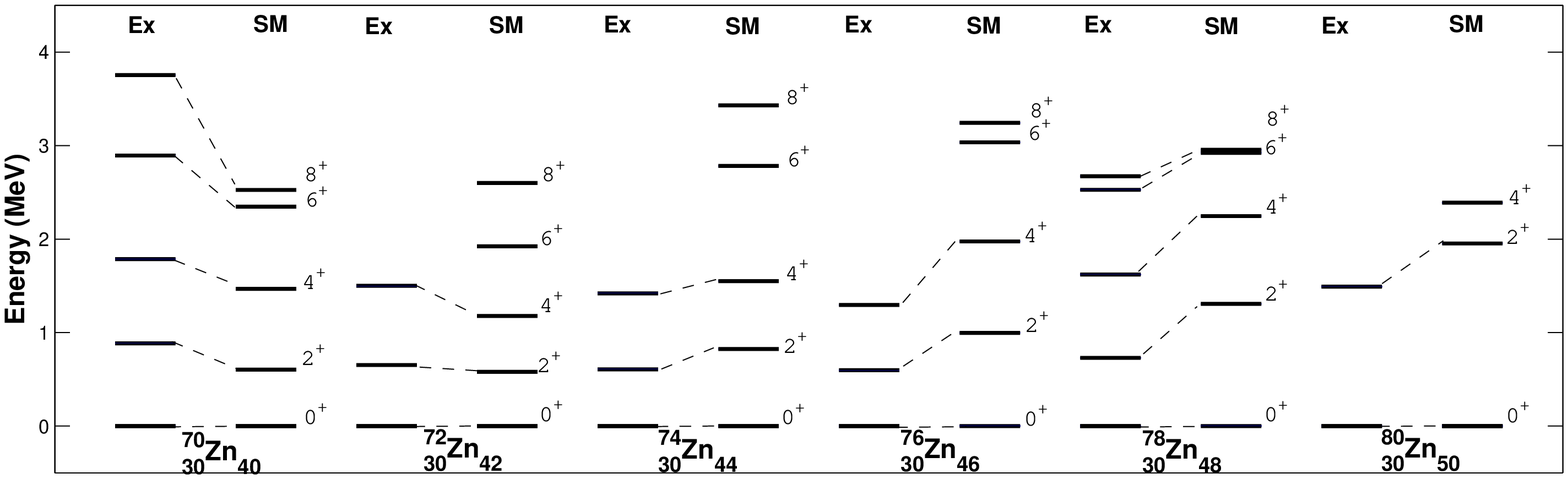}}
\caption{Yrast levels of $^{70-80}$Zn isotopes with
$pfg9a$ interaction.}
\end{center} 
\end{figure}

\begin{figure}[h]
\begin{center}
\resizebox{115mm}{!}{\includegraphics{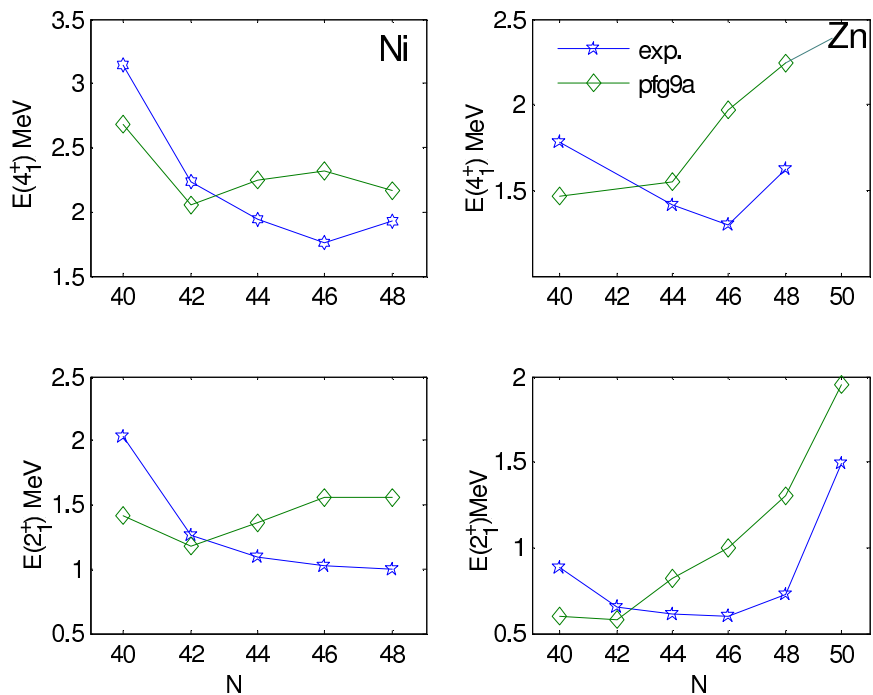}}
\caption{The calculated and experimental E(2$_1^+$)
and E(4$_1^+$) for Ni and Zn isotopes as a function of neutron number.}
\end{center} 
\end{figure}

\subsection{ The $B(E2)$ systematics in the copper isotopes}

The calculated values of first excited states of 1/2$^-$, 5/2$^-$ and 7/2$^-$
with its corresponding experimental values are shown in Fig. 12 for $^{69-79}$Cu
isotopes. In the lower part of this figure, the calculated  and experimental
$B(E2)$ values corresponding to different transitions  namely,
$B(E2; 1/2^-\rightarrow3/2^-)$, $B(E2; 5/2^-\rightarrow3/2^-)$ and
$B(E2; 7/2^-\rightarrow3/2^-)$ are also shown. The calculated and
experimental $B(E2)$ values are tabulated in Table 2. The previously calculated
value by Stefanescu $\it {et ~al.}$ are shown within bracket. \cite{Ste08} The
calculated $B(E2; 1/2^-\rightarrow3/2^-)$ values show much better agreement 
with experimental data compared to those of  Stefanescu. The low $B(E2)$ value
beyond N=40 for 5/2$^-$ state confirms its $\pi$$f_{5/2}$  single-particle
character. The sharp drop in the excitation energy of 5/2$^-$ state beyond N=40 
could be due to the monopole migration. The large $B(E2)$value for 1/2$^-$ state
depart it from the single-particle character of $\pi$$p_{1/2}$ type.
The small $B(E2; 7/2^-\rightarrow3/2^-)$ values for $^{73,75,77,79}$Cu isotopes can be 
seen from wavefunctions of 7/2$^-$ and 3/2$^-$ states in Table 3, which 
show single-particle character of these states. In the first part of the Fig. 12, for the $E(1/2^-)$ state the difference between
 predicted and experimental energy for $^{69}$Cu, $^{71}$Cu, $^{73}$Cu isotopes increases and its
 corresponding $B(E2)$ value also increases. 

\begin{figure}[h] 
\begin{center} 
\resizebox{140mm}{!}{\includegraphics{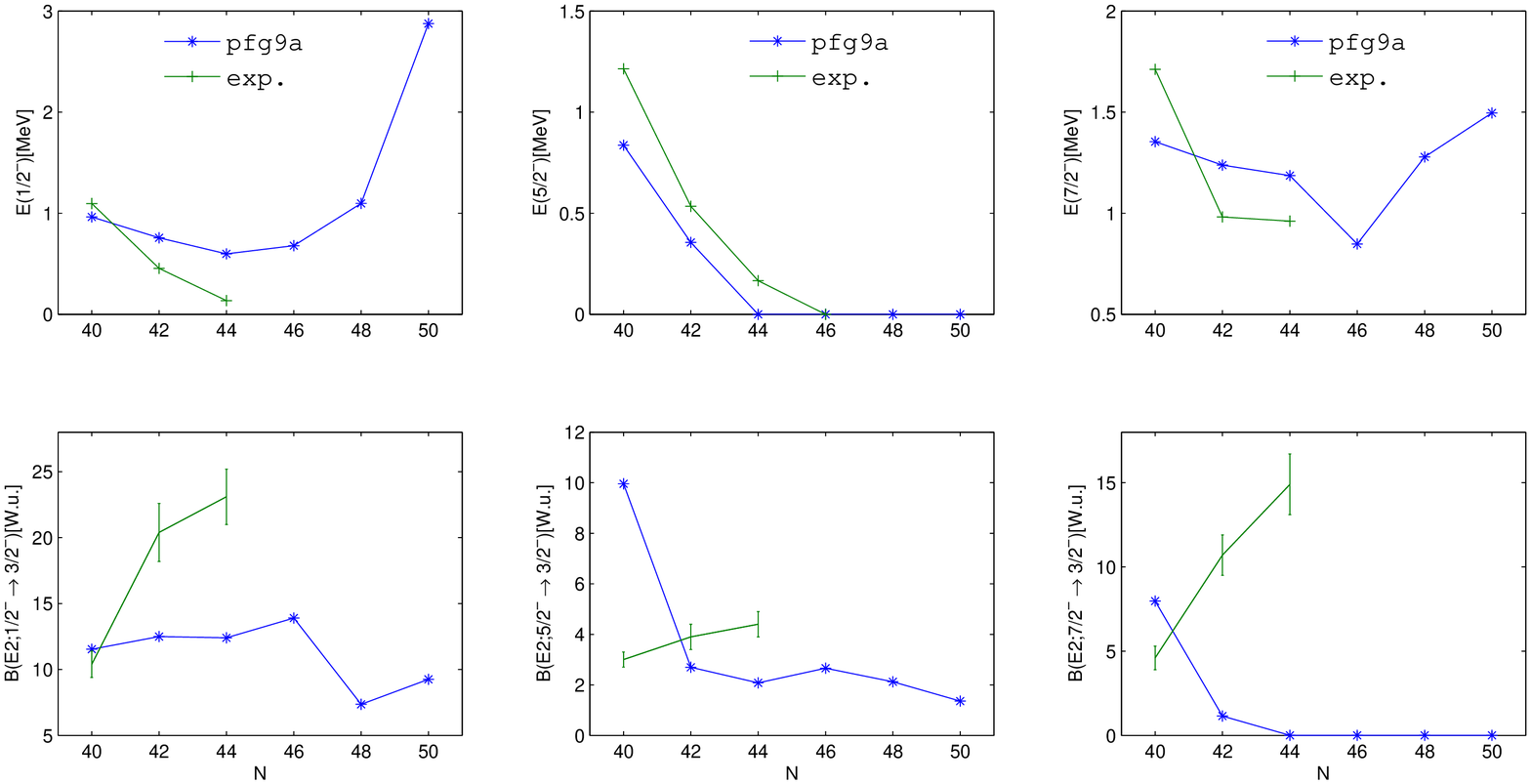}}
\caption{The calculated and experimental $B(E2)$ in W.u. for $^{69-79}$Cu isotopes.
The shell model $B(E2)$ values were calculated with the standard effective charges $e_\pi=1.5e$ and $e_\nu=0.5e$.}
\end{center}
\end{figure} 

\begin{figure}[h]
\begin{center}
\resizebox{60mm}{!}{\includegraphics{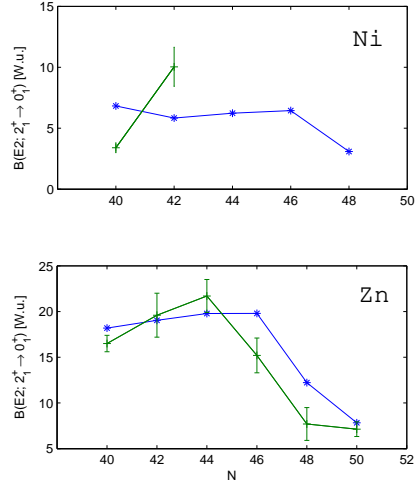}}
\caption{The calculated and experimental
$B(E2; 2_1^+ \rightarrow 0_1^+)$ in W.u. for Ni and Zn isotopes as a function of neutron number.
The shell model $B(E2)$ values were calculated with the standard effective charges $e_\pi=1.5e$ and $e_\nu=0.5e$.}
\end{center}
\end{figure}

\subsection{ The $B(E2)$ systematics in the nickel isotopes} 

The $B(E2; 2_1^+\rightarrow 0_1^+)$ value in $^{68}$Ni is much lower as
compared to $^{56}$Ni. This low  value was interpreted by Sorlin $\it {et ~al.}$
 \cite{Sor02} as originating from the enhanced neutron pair scattering at $N=40$, which
is referred to as superfluid behavior of the neutrons. The constancy of the
$B(E2)$ values beyond $N=40$  can be understood as follows: $\it {E2}$
strength as $\it {np-nh}$ excitation across $N=40$ for odd n cannot
contribute due to parity conservation and for even $\it n$ are dominated by
pair scattering.  Langanke $\it {et ~al.}$  \cite{Lan03} performed microscopic
calculations of the $B(E2)$ in even-even nickel isotopes and found that the
small observed $B(E2)$ value is not necessarily an argument for a shell closure
at N=40, but it simply reflects the fact that the lowest 2$^+$ state in
$^{68}$Ni is primarily a neutron excitation. Van de Walle $\it {et ~al.}$
\cite{Wal07} showed that  $B(E2)$ values in the Ni chain show a parabolic
evolution between two magic numbers (N=28 and 40), hinting a seniority-like
behavior. Perru $\it {et ~al.}$  \cite{Per06} measured high $B(E2)$ value for
$^{70}$Ni at GANIL, and attributed it due to rapid proton core polarization when
neutrons are added to the 1g$_{9/2}$ orbit. The calculated
$B(E2; 2_1^+\rightarrow 0_1^+)$ values for $^{68-76}$Ni are shown in Fig. 13, the
$B(E2)$ value is almost constant from $^{68-74}$Ni and its value again decreases
for $^{76}$Ni. This decrease in $B(E2)$ value is probably due to  oncoming of
the next shell closure at N=50 for $^{78}$Ni.

\subsection{ The $B(E2)$ systematics in the zinc isotopes}

The experimental $B(E2; 2_1^+\rightarrow 0_1^+)$ values in the Zn isotopic
chain show a similar trend towards $N= 40$ as the Ni isotopes up to
$^{68}$Zn. Though for $^{70}$Zn ( at $N= 40$), the $B(E2)$ value suddenly
increases. Leenhardt $\it {et ~al.}$  \cite{Lee02} give three effects for
supporting this increased collectivity: the addition of two protons out of the
Ni core, the maximum in neutron pairing correlations at $N = 40$, and the
presence of the strongly downslopping $l = 4$ Nilsson neutron orbitals
close to the Fermi surface. Kenn $\it {et ~al.}$ \cite{Ken02} indicate that the
inclusion of the $1g_{9/2}$ orbit in the valence space is important in order to
reproduce the increased $B(E2; 2_1^+\rightarrow 0_1^+)$ values in the
$^{70}$Zn. 

 The calculated  $B(E2; 2_1^+\rightarrow 0_1^+)$ values from present work using $pfg9a$ interaction for $^{70-80}$Zn are shown in Fig. 13. The calculated value show similar trends as experimental values from $^{70}$Zn to $^{80}$Zn. The $B(E2)$ value at N=50 is very low, this is an indication of shell closure at this value.

%\newpage
\begin{table}[h]
\tbl{$B(E2)$ values (in W.u.) for $^{69-79}$Cu isotopes. The calculated results shown with corresponding value in the bracket from earlier work of Stefanescu {\it et al.}\protect\cite{Ste08} The shell model $B(E2)$ values were calculated with the standard effective charges $e_\pi=1.5e$ and $e_\nu=0.5e$.}
{\begin{tabular}{@{}ccccccc@{}} \toprule
%\caption{ $B(E2)$ values (in W.u.) for $^{69-79}$Cu isotopes. The results in the bracket from earlier work of Stefanescu.}
%\label{t_bee2}
%\resizebox{14cm}{3.5cm}{
%\begin{ruledtabular}
%\begin{tabular}{rcrcrcr}
          &  $^{69}$Cu$_{40}$ &  $^{71}$Cu$_{42}$  ~~~&   $^{73}$Cu$_{44}$ & $^{75}$Cu$_{46}$ &~ $^{77}$Cu$_{48}$ &   $^{79}$Cu$_{50}$	\\	   
\colrule
           &   Expt. ~~  $pfg9a$ &   Expt. ~~  $pfg9a$ &   Expt. ~~  $pfg9a$ &   Expt. ~~  $pfg9a$ &    ~Expt. ~$pfg9a$ &~  Expt. ~$pfg9a$  \\\\

$B(E2;1/2^-\rightarrow 3/2^-)$ &~~ 10.4[1.0] ~~ 11.5(7.1) &~~ 20.4[2.2] ~~ 12.5(7.3) &~~ 23.1[2.1] ~~ 12.41(7.5) &~~ ~~ 13.9&~~~~7.5&~~~~ 9.2 \\\\

$B(E2;5/2^-\rightarrow 3/2^-)$ &~~ 3.0[0.3]  ~~ 9.9(1.6) &~~3.9[0.5]  ~~2.7(1.7) &~~ 4.4[0.5] ~~ 2.07(1.3)&~~~~ 2.7 &~~~~ 2.1&~~~~ 1.4\\\\

$B(E2;7/2^-\rightarrow 3/2^-)$ &~~ 4.6[0.7]  ~~ 7.9(1.2) &~10.7[1.2] ~~1.2(1.5) &~~ 14.9[1.8] ~~ 0.0038(2.3) &~~~~ 0.0032&~~~~0.0002&~~~~ 0.0016 \\\\
\botrule 
\end{tabular}\label{t_bee2}  }
\end{table}

\begin{landscape}
\begin{table}[h]
\vspace{2.0cm}
\title{Table 3.  The calculated wave-function components of the lowest 5/2$^-$, 3/2$^-$, 7/2$^-$, and  1/2$^-$ states in Cu isotopes.}
\resizebox{19.5cm}{2.0cm}{
\begin{tabular}{ccccc}
\hline
         &  5/2$^-$ &  3/2$^-$  &  7/2$^-$  &  1/2$^-$\\	   
\hline       
$^{69}$Cu$_{40}$        &~~~~  12\% ~~~~$\pi (f_{7/2}^8$$p_{3/2}^1$)$\nu (f_{7/2}^8p_{3/2}^4f_{5/2}^4p_{1/2}^2g_{9/2}^2$)  &~~~~  31\% ~~~~$\pi (f_{7/2}^8$$p_{3/2}^1$)$\nu (f_{7/2}^8p_{3/2}^4f_{5/2}^4p_{1/2}^2g_{9/2}^2$)   &~~~~  22\% ~~~~$\pi (f_{7/2}^8$$f_{5/2}^1$)$\nu (f_{7/2}^8p_{3/2}^4f_{5/2}^4p_{1/2}^2g_{9/2}^2$) &~~~~  16\% ~~~~$\pi (f_{7/2}^8$$p_{1/2}^1$)$\nu (f_{7/2}^8p_{3/2}^4f_{5/2}^4p_{1/2}^2g_{9/2}^2$)\\\\

$^{71}$Cu$_{42}$        &~~~~  18\% ~~~~$\pi (f_{7/2}^8$$f_{5/2}^1$)$\nu (f_{7/2}^8p_{3/2}^4f_{5/2}^4p_{1/2}^2g_{9/2}^4$)  &~~~~  24\% ~~~~$\pi (f_{7/2}^8$$p_{3/2}^1$)$\nu (f_{7/2}^8p_{3/2}^4f_{5/2}^4p_{1/2}^2g_{9/2}^4$)   &~~~~  14\% ~~~~$\pi (f_{7/2}^8$$f_{5/2}^1$)$\nu (f_{7/2}^8p_{3/2}^4f_{5/2}^5p_{1/2}^1g_{9/2}^4$) &~~~~  13.5\% ~~~~$\pi (f_{7/2}^8$$p_{1/2}^1$)$\nu (f_{7/2}^8p_{3/2}^4f_{5/2}^4p_{1/2}^2g_{9/2}^4$)\\\\

$^{73}$Cu$_{44}$         &~~~~  17\% ~~~~$\pi (f_{7/2}^8$$f_{5/2}^1$)$\nu (f_{7/2}^8p_{3/2}^4f_{5/2}^6p_{1/2}^2g_{9/2}^4$)  &~~~~  34\% ~~~~$\pi (f_{7/2}^8$$p_{3/2}^1$)$\nu (f_{7/2}^8p_{3/2}^4f_{5/2}^6p_{1/2}^2g_{9/2}^4$)   &~~~~  15\% ~~~~$\pi (f_{7/2}^8$$f_{5/2}^1$)$\nu (f_{7/2}^8p_{3/2}^4f_{5/2}^5p_{1/2}^1g_{9/2}^6$) &~~~~  11\% ~~~~$\pi (f_{7/2}^8$$p_{1/2}^1$)$\nu (f_{7/2}^8p_{3/2}^4f_{5/2}^6p_{1/2}^2g_{9/2}^4$)\\\\

$^{75}$Cu$_{46}$       &~~~~  35.5\% ~~~~$\pi (f_{7/2}^8$$f_{5/2}^1$)$\nu (f_{7/2}^8p_{3/2}^4f_{5/2}^6p_{1/2}^2g_{9/2}^6$)  &~~~~  36\% ~~~~$\pi (f_{7/2}^8$$p_{3/2}^1$)$\nu (f_{7/2}^8p_{3/2}^4f_{5/2}^6p_{1/2}^2g_{9/2}^6$)   &~~~~  11\% ~~~~$\pi (f_{7/2}^7$$f_{5/2}^2$)$\nu (f_{7/2}^8p_{3/2}^4f_{5/2}^4p_{1/2}^2g_{9/2}^8$) &~~~~  19\% ~~~~$\pi (f_{7/2}^8$$f_{5/2}^1$)$\nu (f_{7/2}^8p_{3/2}^4f_{5/2}^6p_{1/2}^2g_{9/2}^6$)\\\\

$^{77}$Cu$_{48}$      &~~~~  60\% ~~~~$\pi (f_{7/2}^8$$f_{5/2}^1$)$\nu (f_{7/2}^8p_{3/2}^4f_{5/2}^6p_{1/2}^2g_{9/2}^8$)  &~~~~  47\% ~~~~$\pi (f_{7/2}^8$$p_{3/2}^1$)$\nu (f_{7/2}^8p_{3/2}^4f_{5/2}^6p_{1/2}^2g_{9/2}^8$)   &~~~~  41\% ~~~~$\pi (f_{7/2}^7$$f_{5/2}^2$)$\nu (f_{7/2}^8p_{3/2}^4f_{5/2}^6p_{1/2}^2g_{9/2}^8$) &~~~~  50\% ~~~~$\pi (f_{7/2}^8$$f_{5/2}^1$)$\nu (f_{7/2}^8p_{3/2}^4f_{5/2}^6p_{1/2}^2g_{9/2}^8$)\\\\

 $^{79}$Cu$_{50}$      &~~~~  79\% ~~~~$\pi (f_{7/2}^8$$f_{5/2}^1$)$\nu (f_{7/2}^8p_{3/2}^4f_{5/2}^6p_{1/2}^2g_{9/2}^{10}$)  &~~~~  62\% ~~~~$\pi (f_{7/2}^8$$f_{5/2}^1$)$\nu (f_{7/2}^8p_{3/2}^4f_{5/2}^6p_{1/2}^2g_{9/2}^{10}$)   &~~~~  74\% ~~~~$(\pi f_{7/2}^7$$f_{5/2}^2$)$\nu (f_{7/2}^8p_{3/2}^4f_{5/2}^6p_{1/2}^2g_{9/2}^{10}$) &~~~~  57\% ~~~~$\pi (f_{7/2}^8$$p_{1/2}^1$)$\nu (f_{7/2}^8p_{3/2}^4f_{5/2}^6p_{1/2}^2g_{9/2}^{10}$)\\\\

\hline
\end{tabular}}
\end{table}
\end{landscape}

%=================================================================
\section{Conclusion}

In the present work, the existing $pfg9$ interaction for $fpg_{9/2}$
space with $^{40}$Ca core has been modified by changing $\pi f_{5/2}\nu g_{9/2}$ 
 and $\pi f_{7/2}\nu g_{9/2}$  matrix elements. The new interaction is named as $pfg9a$
 has been tuned
for Cu isotopes and tested for Ni and Zn isotopes. The $pfg9a$ interaction
give $0^+$ ground state in an even-even nucleus like in Ni and Zn isotopes which
is the characteristic of any reasonable interaction. These results indicate that
further modification in interaction and even inclusion of $1d_{5/2}$ orbit is
important in the shell model calculations. The modification of Sorlin {\it et.
al.,} interaction is also attempted in,\cite{Sieja10} but it will be remain to
test this interaction is universal or not for this region. An attempt by
including $1d_{5/2}$ orbit in $fpg_{9/2}$ space to explain collectivity for $fp$ 
shell nuclei is recently reported in. \cite{Lenzi10}

%-------------------------------------------------------------------------

\section*{Acknowledgments}

Thanks are due to Prof. P.~Van Isacker and  Prof. I.~Mehrotra for suggestions during the work. All
the calculations in the present paper are carried out using the SGI cluster
resource at GANIL and DGCTIC-UNAM computational facility KanBalam. This work was supported in part by
grants from the Sandwich PhD
programme of the Embassy of France in India, Conacyt, M\'exico, and by DGAPA, UNAM project IN103212.
I acknowledge fruitful discussions with Dr. 
M. J. Ermamatov. The discussions with Prof. J. G. Hirsch and Prof. O. Civitarese is also acknowledge.

%\newpage
%-------------------------------------------------------------------------

%------------------------------------------------------------------

\end{document}